# Geometrical determinant of nonlinear synaptic integration in human cortical pyramidal neurons

Jaeyoung Yoon [1,2,*]

[1] McGovern Institute for Brain Research, Massachusetts Institute of Technology, Cambridge, MA 02139, USA.
[2] Present address: F. M. Kirby Neurobiology Center, Boston Children's Hospital, Harvard Medical School, Boston, MA 02115, USA.
[*] Correspondence: jy.yoon@tch.harvard.edu

**Abstract**

Neurons integrate synaptic inputs and convert them to action potential output at electrically distant locations. The computational power of a neuron is hence enhanced by subcellular compartmentalization and nonlinear synaptic integration, but the biophysical determinants of these features in human neurons are not completely understood. By examining the synaptic input-output function of human neocortical pyramidal neurons, we found that the nonlinearity threshold at the soma was linearly determined by the shortest path distance from the synapse to the apical trunk, and the slope of this relationship was consistent throughout the dendritic arbor. Analogous rules were found from both supragranular and infragranular layers of the rodent cortex, suggesting that these represent a fundamental property of pyramidal neurons. Additionally, we found that neurons associated with tumor or epilepsy had distinct membrane properties, but the nonlinearity threshold was shifted in amplitude such that the slope of its relationship with synaptic distance remained consistent.

**Introduction**

The cortical pyramidal neuron (PN) transforms synaptic inputs at its dendrites into action potential (AP) output near the soma. As the input and the output structures of the neuron are electrically distant, the transfer function of the neuron is determined jointly by the cable properties of the dendritic arbor and the distribution of ionic conductances, in addition to the spatiotemporal pattern of synaptic inputs. [1, 2] Functional compartmentalization in dendrites can thus enable each neuron to perform more complex operations by utilizing electrical or chemical mechanisms, [3, 4, 5, 6] in such way that the single neuron can resemble multiple layers of processing units rather than a simple point neuron. [7, 8, 9, 10] A crucial component of this feature is nonlinear synaptic integration, which is mediated primarily by the voltage-dependent activation of N-methyl-D-aspartate receptors (NMDAR) from synchronized inputs, [11, 12] facilitated by the opening of α-amino-3-hydroxy-5-methyl-4-isoxazolepropionic acid receptors (AMPAR) that are relatively voltage-independent. These synaptic conductances can interact further with other regenerative mechanisms such as the recruitment of voltage-dependent $Ca^{2+}$ or $Na^+$ conductances in generating dendritic spikes, [13, 14, 15, 16] which have been implicated in functions such as coincidence detection, burst firing, plasticity, and stimulus selectivity. [17, 18, 19, 20, 21]

It has been shown recently that human layer 2/3 (L2/3) PNs are capable of NMDAR-mediated nonlinear synaptic integration, similar to what had been characterized from rodent neurons in literature. [22] Nevertheless, a quantitative description of the biophysical factors that determine the engagement and the extent of nonlinear synaptic integration in human neurons based on empirical evidence has not been established. In the previous study which used glutamate iontophoresis or extracellular electric stimulation, only a small fraction of human L2/3 PNs (< 30%) displayed NMDAR-mediated nonlinearity, along with highly variable somatic excitatory postsynaptic potential (EPSP) amplitudes; on the contrary, the majority of mouse L2/3 PNs (> 70%) showed NMDAR spikes with smaller variance in amplitude. Notably, such an apparent contrast between human and mouse neurons was observed despite the overall lack of difference in the synaptic AMPAR:NMDAR ratio measured from the same cells. Based on simulations from a model neuron, it was further hypothesized that the presumed low propensity for NMDAR spikes in human L2/3 PNs could be explained by the higher local synaptic conductance threshold, which was mainly attributed to larger dendritic diameter amongst other passive properties that were explored as variables. [22] As with any model, these conclusions depend critically on the inherent assumptions and parameters, thus it is essential that they are compared with sufficient experimental evidence. However, empirical knowledge on the cellular physiology of human neurons has been particularly limited due to the difficulties associated with using human brain tissue for ex vivo electrophysiology.

In the present study, we examined the synaptic integration properties of human neurons with whole-cell patch clamp from neocortical L2/3 PNs in acute brain slices obtained from adult patients, combined with 2-photon glutamate uncaging (2PGU) and calcium imaging to activate synaptic spines with spatial and temporal precision while monitoring local nonlinear events. We conclude that all human L2/3 PN dendrites are equally capable of supralinear synaptic integration, wherein the nonlinearity threshold at the soma is linearly determined by the shortest path distance from the synapse cluster to the apical trunk. The slope of this linear relationship was consistent for synapses throughout the dendritic arbor, and analogous rules were found from both L2/3 and L5 PNs in the rodent cortex. In addition, we found that human L2/3 PNs associated with heterogenous tumor or epilepsy had distinct intrinsic membrane properties, even though in either case they originated from areas that were clinically categorized as nonpathological. Intriguingly, L2/3 PNs in these parts of the cortex from epilepsy patients had lower intrinsic excitability compared to those from tumor patients, as well as higher nonlinearity threshold across synaptic distances, which was reflected as a shift in the intercept but not the slope of their linear relationship.

**Results**

*Human neuronal dendrites are equally capable of nonlinear synaptic integration*

We investigated the synaptic integration properties of human neurons by characterizing the synaptic input-output function of neocortical L2/3 PNs with whole-cell patch clamp at the soma while activating synaptic spines along a same dendritic branch segment by 2PGU at diverse locations within the dendritic arbor, simultaneously with intracellular calcium imaging at the parent dendrite of the activated spines. Recordings were made in acute human brain slices (n = 172 cells, 22 humans; **Table 1**), prepared from parts of the neocortex that were considered to be nonpathological from clinical evaluations but resected during the surgical procedure. Cortical tissue originated from the temporal and frontal lobes of adult patients diagnosed with tumor (n = 43 cells, 5 humans) or epilepsy (n = 129 cells, 17 humans). Cells were identified as L2/3 PNs based on their intrinsic electrophysiological properties as well as anatomy visualized with gradient contrast and 2-photon excitation microscopy, including somatodendritic morphology, presence of dendritic spines, and cortical depth (**Fig. 1a**).

First, we performed 2PGU at synapses located on the apical tuft dendrites, which were at considerable distances from the soma in human L2/3 PNs (**Fig. 1b-c**). Larger responses from simultaneous activation of multiple synaptic spines were associated with the typical characteristics of NMDAR conductance, namely the slow channel kinetics manifested on the time course of the EPSP waveform coincident with postsynaptic $Ca^{2+}$ influx (**Fig. 1d**). Consequently, the measured EPSP at this point exceeded what would be expected from a hypothetical scenario with an entirely linear synaptic integration scheme, wherein this latter expected EPSP was calculated as the arithmetic sum of unitary EPSPs (uEPSP) that were measured independently from respective single spines (**Fig. 1e**). Synaptic gain, defined as the amplitude ratio of the measured vs. expected EPSP, was therefore above unity in this regime, indicative of supralinear integration (**Fig. 1f**). Since uEPSP amplitudes varied across spines, the threshold for nonlinearity was defined for each synapse cluster on a given dendritic branch segment as the point at which $Ca^{2+}$ signal was first observed from the parent dendrite of the activated spines. For grouping purposes, the synaptic gain and the fractional change in $Ca^{2+}$ fluorescence (dF/F) were then aligned to this nonlinearity threshold in terms of the number of activated spines (**Fig. 1g**). Similar to the tuft dendrites, both the oblique dendrites branching from the apical trunk (**Fig. 1h-l**) and the basal dendrites extending from the soma (**Fig. 1m-q**) likewise exhibited supralinear integration with the typical characteristics of NMDAR activation, as had been confirmed previously. [22]

It remains to be clarified whether the nonlinear effect at the soma produced by distal apical synapse activation was mediated primarily by passive propagation of the EPSP, or possibly with additional involvement of other active conductances at intermediate locations along the apical trunk as has been found in the case of L5 extratelencephalic (ET) neurons. [15, 23] Regardless, the presence of the postsynaptic $Ca^{2+}$ signal and the associated EPSP kinetics indicated that these events were initiated at the synapse via NMDAR activation. [22] Synaptic gain did not always increase monotonically with the number of activated spines due to the saturation of synaptic membrane potential, [16] which was supported by the continued occurrence of postsynaptic $Ca^{2+}$ influx in these situations. Relatedly, activation of distal tuft synapses along a single dendritic branch segment was often insufficient to generate somatic APs in human L2/3 PNs with up to 35 spines activated simultaneously, even with supralinear integration. As an additional note, we did not observe an anticoincidence (XOR) logic from the synaptic input-output functions of human L2/3 PNs by activating apical tuft synapses with 2PGU while recording from the soma, unlike what was reported from a previous study in the form of decreasing peak amplitudes of dendritic $Ca^{2+}$-mediated action potentials (dCaAP) in response to increased direct current injection at the recording site near the nexus. [24]

*Somatic nonlinearity threshold is linearly determined by synaptic distance to the apical trunk*

Although we had initially grouped synapses according to a conventional categorization for their dendritic location (basal, apical oblique, and apical tuft), the precise position of these synapses nonetheless varied not only across groups but also within each group, such as regarding the distance to the apical trunk or the soma, relative position along the dendritic branch segment, branch order, or depth within the cortical column. Similarly, while most synapses were capable of nonlinear integration, heterogeneity in the shapes of the synaptic input-output functions was present such as in terms of the nonlinearity threshold, synaptic gain, or the degree of saturation. Hence, we inspected the electrophysiological parameters constituting these input-output curves with respect to the associated anatomical parameters (**Fig. 2a-c**). We conclude that the somatic depolarization at supralinearity threshold, henceforth denoted as $V_{ST}$ (**Fig. 2c**), is linearly determined by the path distance along the dendrite from the synapse cluster to the nearest point on the apical trunk (**Fig. 2d-g**). $V_{ST}$ was defined as the amplitude of the expected EPSP calculated from the sum of uEPSPs at respective spines, upon simultaneous activation of which nonlinearity was first observed from the postsynaptic response as evidenced by the $Ca^{2+}$ signal at the parent dendrite. For the definition of $V_{ST}$, the expected EPSP from the linear sum of uEPSPs was used instead of the EPSP measured from simultaneous activation of the same spines, to correctly determine the nonlinearity threshold independent of the extent of nonlinearity, i.e. the nonlinear component of the EPSP. Since we defined the position of the soma as the center of mass of its boundary, path distance from basal dendrite synapses to the apical trunk or the soma are loosely synonymous as the difference is only in the order of a few micrometers. It also follows that for tuft dendrite synapses, the synaptic distance defined in this context is equal to their distance to the nexus.

The nonlinearity threshold, $V_{ST}$, had a strong linear anticorrelation with synaptic distance to the apical trunk, either separately for basal, oblique, and tuft dendrites (**Fig. 2d-f**), or when all synapses were considered together (**Fig. 2g**). Furthermore, the slope of the linear regression was consistent across all groups of dendrites (F = 0.06380, P = 0.9383). In contrast, $V_{ST}$ could not be adequately explained using the distance to the soma instead (**Fig. 2h-k**). For clarity, we stress that $V_{ST}$ represents the somatic membrane depolarization at which point the local membrane potential near the synapse reached the nonlinearity threshold. One possible interpretation of these results could be that, in human L2/3 PNs, the length constant for EPSPs in the dendrosomatic direction is similar for all thin dendritic branches while attenuation along the apical trunk is relatively minor. Nevertheless, there is insufficient evidence at this time regarding the precise synaptic depolarization or the degree of dendritic filtering for synaptic events in human neurons, such that a mechanistic analysis of these properties requires further investigation. From rodent neurons, it had been shown that the generation of NMDAR spikes produces larger synaptic gain and broader window for temporal summation at relatively more distal thirds of their basal and oblique dendrites, tested with a small number (≤ 7) of spines. [25] By exploring a wider range of synaptic inputs (≤ 35 spines) with physiologically relevant uEPSP amplitudes for human neurons, [26] and using an absolute rather than a relative metric of distance, we did not find a quantitative relationship for maximal synaptic gain with respect to synaptic distance either from the apical trunk or the soma (**Fig. 2l-m**), nor a categorical difference according to dendrite type (**Fig. 2n**). On the other hand, $V_{ST}$ tended to be larger for tuft dendrites compared to basal or oblique dendrites (**Fig. 2o**), but other factors were more likely responsible for this difference (discussed below).

Due to practical reasons associated with the difficulties of using human cortical tissue for acute brain slice experiments, the scarce body of literature including ex vivo electrophysiology in human neurons have often made use of tissue obtained from epilepsy patients due to its availability, and sometimes additionally from tumor patients as a control. Although many of these works, including the present study, have used areas that were clinically categorized as nonpathological, the lack of empirical knowledge at the cellular resolution from the human brain in healthy conditions poses an insurmountable challenge to the detailed understanding of human neurophysiology. The issue is further complicated by the fact that epilepsy patients are often chronically treated with antiepileptic drugs (AED) which are designed to interfere with the very same fundamental properties of neurons that are the subject matter of the study. Tissue associated with tumor is neither free of similar concerns, given that tumors are known to affect the synaptic properties of adjacent areas, [27, 28] the spatial extent of which is ambiguous and cannot be delineated easily. In order to identify possible systematic differences in the synaptic integration properties associated with tissue origin, we reanalyzed our data by grouping them according to the diagnosis of the patient. Intriguingly, we found that L2/3 PNs from neocortical tissue obtained from epilepsy patients had consistently higher $V_{ST}$ compared to those from the tumor group (**Fig. 2p-s**). By chance, all but one data from apical tuft dendritic locations more distal to the nexus were collected from the epilepsy group, but also with the single exception from the tumor group coinciding with the lowest $V_{ST}$.

*Distinct intrinsic membrane properties of human neurons associated with tumor or epilepsy*

Since there was a noticeable divergence in terms of the nonlinearity threshold of L2/3 PNs according to tissue origin, we further inspected the intrinsic membrane properties of neurons from cortices associated with tumor or epilepsy. Previously, we have shown that both the intrinsic and the synaptic properties of human L2/3 PNs and fast-spiking interneurons (FSIN) are distinct between the non-epileptic and the epileptic cortex, with emphasis on the net synaptic drive of FSINs being inverted from excitation to inhibition to represent the mechanism responsible for network hyperactivity in epilepsy associated with focal cortical dysplasia type I (FCD I). [29] As an important note, these previous comparisons were made between the nonpathological part of the cortex associated with heterogenous tumor versus the pathological area within the ictal onset zone of the epileptic cortex, whereas the present study used cortical tissue from tumor or epilepsy patients that was in either case considered to be unaffected by their respective pathophysiology, based on clinical evaluations prior to and following surgery. The initial assumption was therefore to find no difference in the basic neuronal properties between these groups. Nevertheless, data indicated a clear division between the intrinsic membrane properties of L2/3 PNs, also in a possibly counterintuitive direction that neurons from epilepsy had lower intrinsic excitability compared to the tumor group (**Fig. 3a**). On average, L2/3 PNs from epilepsy had lower input resistance ($R_{in}$) (**Fig. 3b**), accompanied by higher rheobase (**Fig. 3c**), compared to L2/3 PNs from tumor. These differences appeared to be in the opposite direction from said previous comparisons between tumor and epilepsy, where both $R_{in}$ and firing frequency of L2/3 PNs were higher in epilepsy; but importantly from L2/3 PNs within the ictal onset zone instead of a nonpathological area. [29] The lower $R_{in}$ of L2/3 PNs from epilepsy in the current study was despite their also having smaller hyperpolarization-activated and cyclic-nucleotide-gated (HCN) channel conductance, represented by the lower sag ratio (**Fig. 3d**). Resting membrane potential (RMP), on the other hand, was not significantly different between groups (**Fig. 3e**). Additionally, we note that similar differences could be noticed from a previous dataset, even though the distinction between tumor and epilepsy was not made for its interpretation. [30]

It has been reported that both the hyperpolarization-activated current ($i_h$) through HCN channels and the $R_{in}$ of human L2/3 PNs form a gradient along the cortical depth, with a tendency for larger $i_h$ and lower $R_{in}$ and for cells deeper within L2/3. [30, 31] At least some of the difference in the $R_{in}$ between L2/3 PNs from tumor or epilepsy in the present dataset might be explained by this factor, as there was a bias in the depths of recorded cells (**Fig. 3f**); however, the difference in the sag ratio between groups was the opposite of what would be expected from this correlation alone. Moreover, the lower $R_{in}$ and sag ratio of L2/3 PNs in epilepsy compared to tumor were still conspicuous even when these properties were compared across cortical depths (**Fig. 3g-h**). In either group, there was little variance in the RMP with no significant correlation with cortical depth (**Fig. 3i**), contrary to the previous report. [30] These may have been simply related to the distribution of cortical depths of cells included in the current dataset, with more than 78% of all recorded cells positioned within 500-1000 μm (cortical depth, 862 ± 29 (μm); $P_{90}$ = 1137, $P_{10}$ = 572; κ = 0.52). We also note that combining data indiscriminative of tumor or epilepsy similarly to previous works results in nominally stronger statistical significance for correlations from this larger collective sample, for both $R_{in}$ (ρ = -0.4154, P < 0.001) and $i_h$ (ρ = +0.3379, P < 0.001) with respect to cortical depth. In addition, $R_{in}$ or the sag ratio can represent the underlying ionic conductances to different degrees according to the calculation method, for which unfortunately there is no consensus (see Methods). Here, we used a linear regression crossing the origin obtained from the transient-state subthreshold membrane potential responses to step current injections, to minimize the contribution of $i_h$ to the calculated values of $R_{in}$.

In agreement with the $R_{in}$, maximum firing rates of L2/3 PNs were either weakly or not significantly correlated with cortical depth, but distinctly lower in epilepsy compared to tumor across varying depths (**Fig. 3j**). Indeed, firing rates of L2/3 PNs from epilepsy were lower for all depolarizing somatic current injection amplitudes (**Fig. 3k**), and maximum firing rates had a strong linear correlation with $R_{in}$ (**Fig. 3l**). A modest difference in spike frequency adaptation was present between groups in terms of the inter-spike interval (ISI) at rheobase (**Fig. 3m**), which was diminished at larger current injection amplitudes (**Fig. 3n**); however, the number of spikes generated at the respective rheobase for each cell or its multiple were lower in epilepsy by nearly twofold (tumor vs. epilepsy; 3.9 ± 0.7 vs. 2.1 ± 0.2 (spikes), P < 0.01 at rheobase; 4.2 ± 0.7 vs. 2.1 ± 0.2, P < 0.001 at 2*rheobase). For a more detailed inspection of the spike properties of human L2/3 PNs, we analyzed the single AP kinetics from the first AP generated at rheobase (**Fig. 3o-p**). AP threshold, defined as the membrane potential at which its derivative with respect to time ($dV_m/dt$) reached 10 (V/s), was not different between tumor and epilepsy (**Fig. 3q**). AP amplitude, calculated as the difference between the AP peak and threshold, was likewise similar (**Fig. 3r**). AP half-width was significantly longer in L2/3 PNs from epilepsy, by approximately 15% on average (**Fig. 3s**). The maximum rate of depolarization was similar between groups (**Fig. 3t**), in alignment with the similar AP threshold, peak, and amplitude. The maximum rate of repolarization was considerably slower in L2/3 PNs from epilepsy (**Fig. 3u**), explaining the longer AP half-width (**Fig. 3v**). Of note, the distribution of the maximum rate of repolarization in L2/3 PNs from epilepsy was substantially skewed, in contrast to the tumor group where it was close to normal distribution (tumor vs. epilepsy; $G_1$, 0.31 vs. 1.73; κ, -0.16 vs. 4.97; see Methods); while further examinations would be required, smaller high-voltage activated $K^+$ channel conductance in L2/3 PNs associated with epilepsy may represent a possible candidate mechanism underlying this difference. [32]

*Analogous rules for nonlinearity threshold across cortical laminae and species*

The somatic depolarization at supralinearity threshold ($V_{ST}$) was determined linearly by the shortest path distance from the synapse to the apical trunk, in all synaptic locations for L2/3 PNs in the human cortex (**Fig. 2**). In order to find whether this rule could be generalized across neuronal subtypes or species, we extended our investigation to the rodent temporal association cortex (TeA) using rat L2/3 and L5 ET PNs, the latter of which are comparable in size with human L2/3 PNs (**Fig. 4a**). Similar to previous experiments, synapses at variable locations were activated by 2PGU to compare $V_{ST}$ with synaptic distance (**Fig. 4b**). With a sufficient number of spines activated simultaneously, the basal and the oblique dendrites of rat PNs produced NMDAR-mediated nonlinearity (**Fig. 4c**), upon which the measured EPSP exceeded the hypothetically expected EPSP from the linear sum of uEPSPs (**Fig. 4d**). Consistent with the previous observations from human neurons, $V_{ST}$ in rat L5 ET cells was again linearly and inversely correlated with synaptic distance from the apical trunk, along with similar slopes for both basal and oblique dendrites (**Fig. 4e-g**). Rat L5 ET cells were therefore capable of supralinear integration in both locations (**Fig. 4h-i**), as had been shown abundantly in the case of rodent neurons. L2/3 PNs followed the same pattern, with $V_{ST}$ linearly anticorrelated with synaptic distance from the apical trunk (**Fig. 4j-n**). These results suggest that analogous rules for dendritic processing are present for PNs across different cortical layers and species. [33]

**Discussion**

We demonstrated that the nonlinearity threshold at the soma is linearly determined by synaptic distance from the apical trunk in both human and rodent PNs (**Fig. 2, 4**). As a corollary, all synapses distributed along the dendritic arbor of these neurons were similarly capable of supralinear integration, so long as they were sufficiently distant from the apical trunk (**Fig. 1**). Some previous works, however, have suggested possible categorical differences in the integration schemes associated with dendrite type to argue that specifically their apical oblique dendrites, but not the basal or the tuft dendrites, may operate predominantly in a linear integration mode in the mature cortex, based on somatic recordings from mouse L5 ET cells in areas involved in visual information processing. [34, 35] Such apparent segregation of synaptic integration rules across subcellular locations was subsequently linked with long-range connectivity or cortical development, with connotations of specialized functions corresponding to the identity of presynaptic input; further, it was proposed that changes in the synaptic AMPAR:NMDAR ratio may be responsible for this phenotype.

While a patterned heterogeneity in the subcellular distribution of synaptic AMPAR:NMDAR ratio remains a possible feature of cortical PNs, [36] we found that the input-output function from synapse clusters in any given location can nonetheless be explained by a simple geometrical factor. From this perspective, if a certain group of synapses were only capable of linear integration, this would be a straightforward consequence of its not being sufficiently distant from the apical trunk such that in this case the nonlinearity threshold $V_{ST}$ would be higher than the somatic AP threshold; in other words, synaptic activation at this location cannot be expected to generate nonlinearity without triggering a somatic AP, which would mask potential nonlinear effects at the soma. One of the previous works using mouse neurons tried to indirectly address this issue by either hyperpolarizing the soma with direct current injection or by bath application of the Na$^+$ channel blocker tetrodotoxin (TTX) to respectively shunt or block AP generation while recording from the soma; [34] however, these results cannot be interpreted unambiguously without quantitative information of the anatomical location of these synapses, especially given that the mouse dendrites are shorter and the uncaging sites in those works which suggested linear integration at the oblique dendrites were as close as 50 μm or more proximal to the apical trunk for those synaptic locations. [34, 35]

All of the electrophysiological recordings in the present study were likewise made from the soma, as whole-cell patch clamp directly at the thin, higher-order dendritic branches is extremely challenging and made even more unrealistic with the limited availability of human tissue. Since the precise synaptic depolarization and the degree of dendritic filtering along the thin dendrites of human neurons are unknown, it is difficult at this point to further dissociate the mechanisms underlying the shapes of the input-output curves, such as in terms of the synaptic AMPAR:NMDAR ratio or the passive and active conductances encountered as the EPSP travels along the dendrite from the synaptic spines to the soma. Moreover, although our experiments were performed with 2PGU calibrated to produce uEPSPs within a physiological range of somatic amplitudes and kinetics known from literature, [26] the synaptic parameters associated with vesicle release events from presynaptic terminals specifically apposed to postsynaptic spines at different subcellular locations of human cortical neurons are unknown, even though they are relevant factors to postsynaptic computation; [37] not to mention those in normal conditions unassociated with any brain lesion or patient medication history. While for these reasons a mechanistic analysis of the geometric rule for nonlinearity threshold herein described requires further clarification, this relationship can be understood with regard to how synaptic inputs at various subcellular locations are processed as they are transformed ultimately to neuronal output. A potential benefit of this property could be that it does not necessarily require a specialized transcription, translation, or trafficking strategy for different cellular subcompartments in order to produce a continuous range of responses corresponding to a range of synaptic inputs.

Understandably due to the scarce availability of human brain tissue suitable for slice electrophysiology, there has been a tendency in practice of collecting data indiscriminately from cortical tissue associated with different pathologies, provided that the parts used for experiments were clinically categorized as nonpathological. [23, 26, 30, 38, 39, 40] However, we found that both the intrinsic and the synaptic properties of neurons associated with heterogenous tumor or epilepsy were consistently distinct (**Fig. 3**), despite no obvious signs of abnormal physiology in the cortical tissue from which recordings were made. These results were not entirely surprising given that in many cases these epilepsy patients had undergone pharmacological treatments and often for extensive periods; such differences in cellular properties could also be compared with a previous study in which contrasting results were found using the pathological area of the epileptic cortex instead. [29] In spite of having made the distinction between tumor and epilepsy, the present work is still not entirely exempt from the caveats of heterogeneity in the human samples. Our data were obtained from a diverse range of conditions in terms of cortical area, hemisphere, patient age, sex, ethnicity, medication history, and other unknown factors of possible relevance; for realistic reasons, these parameters could not be as strictly controlled as in non-human animal experiments.

## Methods

*Brain slice preparation*

All protocols were approved by the respective internal review boards (IRB) of Massachusetts General Hospital (MGH), Brigham and Women's Hospital (BWH), and the institutional animal care and use committee (IACUC) of Massachusetts Institute of Technology (MIT). For acute human brain slices, tissue resected from patients was immediately placed in an ice-cold solution at the operating theater. The cutting solution used for transport and slicing contained (in mM): 165 sucrose, 20 HEPES, 25 $NaHCO_3$, 2.5 KCl, 1.25 $NaH_2PO_4$, 20 D-glucose, 5 sodium ascorbate, 3 sodium pyruvate, 0.5 $CaCl_2$, 7 $MgCl_2$, pH adjusted to 7.3 with NaOH. All materials were obtained from Sigma-Aldrich unless otherwise specified. Tissue was transported from the operating room at MGH or BWH to the laboratory at MIT in a thermally insulated container filled with ice packs, within 25 minutes. Tissue was then placed in a vibratome (VT1200S, Leica) for slicing, orthogonal to the pial surface to preserve all cortical layers up to white matter in proper orientation. Acute brain slices (300 μm in thickness) were made, then maintained at 36 °C for ≥ 1 h in the recovery solution containing (in mM): 90 NaCl, 20 HEPES, 25 $NaHCO_3$, 2.5 KCl, 1.25 $NaH_2PO_4$, 20 D-glucose, 5 sodium ascorbate, 3 sodium pyruvate, 1 $CaCl_2$, 4 $MgCl_2$, pH adjusted to 7.3 with NaOH, 300-310 mOsm. All solutions were continuously aerated with carbogen (95% $O_2$, 5% $CO_2$) throughout the course of experiments, including transport and slicing, and refreshed every 6-8 hours. Experiments were performed within a period of typically ~48 h following resection, but occasionally ~24 or ~72 h, and up to 120 h (**Table 1**). For rodent brain slices, 12-to-13-week-old male and female rats (Charles River) were anesthetized with isoflurane (5% v/v) prior to decapitation; the cutting solution contained (in mM): 225 sucrose, 25 $NaHCO_3$, 2.5 KCl, 1.25 $NaH_2PO_4$, 10 D-glucose, 0.5 $CaCl_2$, 7 $MgCl_2$, and slices were recovered in artificial cerebrospinal fluid (aCSF) identical to the recording solution. A subset of the human slices prepared for the current study was donated to be used elsewhere in an uncredited work. [41]

*Whole-cell patch clamp*

All human and rodent brain slice experiments were conducted under identical conditions. Slices were placed in a recording chamber and visualized with Dodt gradient contrast microscopy (Examiner Z1, Zeiss), while being perfused with the recording aCSF at 1.6 mL/min using a peristaltic pump (Minipuls 3, Gilson). The recording solution contained (in mM): 125 NaCl, 25 $NaHCO_3$, 3 KCl, 1.25 $NaH_2PO_4$, 10 D-glucose, 1.2 $CaCl_2$, 1.2 $MgCl_2$, 300-310 mOsm, maintained at 36 °C with an inline heating system (TC344C & SH28, Warner). Whole-cell patch clamp recordings were made from the soma of pyramidal neurons (PN) in layer 2/3 (L2/3) of the human neocortex, or L2/3 and L5 of the rat temporal association cortex (TeA), using a MultiClamp 700B amplifier (Molecular Devices). Data were acquired at the sampling rate of 20 kHz, bessel filtered at the cutoff frequency of 10 kHz, digitized, and recorded with Prairie View (Bruker), then visualized and analyzed with PVBS (https://github.com/flosfor/pvbs). Patch pipettes (2.5-4.0 MΩ) were pulled from borosilicate glass capillaries (PG52151-4, WPI) with a pipette puller (P-1000, Sutter), and positioned using a set of micromanipulators (Junior, Luigs & Neumann). The internal solution contained (in mM): 130 K-gluconate, 4 KCl, 4 NaCl, 10 HEPES, 15 phosphocreatine-di(tris), 4 Mg-ATP, 0.3 Na$_2$-GTP, adjusted to pH 7.3 with KOH, leading to a final [K$^+$] of ~136-138 mM, and 300-310 mOsm. Additionally, 0.05 mM Alexa Fluor 594 and 0.1 mM Oregon Green BAPTA-1 (Thermo Fisher) were added to the internal solution for structural and calcium imaging. Pipette capacitance ($C_p$) was fully compensated, typically ~13.5-14.5 pF. Series resistance ($R_s$) was continuously monitored throughout the course of recordings, and cells with Rs ≤ 20 MΩ and change of less than 15% were accepted. To measure the intrinsic neuronal membrane properties, cells were first held in current clamp at $i_{cmd}$ = 0 pA, then injected at the soma with square current steps (-250 to +1000 pA in 50 pA steps, duration 500 or 1000 ms), with bridge fully balanced. Previous works in literature have used different methods for calculating the input resistance ($R_{in}$) and the hyperpolarization-activated current ($i_h$), e.g. regarding whether to use a fixed current injection amplitude and in which case what amplitude, [42] or to use variable current amplitudes producing similar membrane potential changes in different cells, [43] or alternatively to use a regression from a range of inputs and responses, [44] and also whether to use the steady-state [45] or the transient [46] membrane potential response. In the present work, $R_{in}$ was calculated from the linear regression crossing the origin from all subthreshold membrane potential responses to step current inputs; for hyperpolarizing pulses, transient-state membrane potential responses were used to minimize the contribution of $i_h$ in calculating the $R_{in}$. Sag ratio, as a measure of $i_h$, was defined as the difference between the transient-state and the steady-state membrane potential responses to -250 pA somatic current injection divided by the former, in alignment with the definition of $R_{in}$. Transient-state responses were calculated as the peak of the response, whereas steady-state responses were calculated as the mean from the later 20% of the step duration. Resting membrane potential (RMP) was calculated as the mean of the membrane potential during the baseline period preceding the current steps, typically 500 ms. For action potential (AP) waveform analysis, the first AP generated at rheobase was used. AP threshold was defined as the membrane potential immediately prior to the point at which the derivative of the membrane potential with respect to time ($dV_m/dt$) first exceeded 10 (V/s); for this purpose, data were further processed with a 4$^{th}$ order bessel filter at the cutoff frequency of 4 kHz for enhanced signal-to-noise ratio. AP amplitude was calculated as the difference between the AP peak amplitude and the AP threshold, and the time elapsed between the two points at half of this amplitude respectively from the depolarizing and the repolarizing phase was taken as AP half-width. Liquid junction potential, measured at ~11.5 mV, was not corrected for in the values of membrane potential reported in the current study.

*2-photon glutamate uncaging and calcium imaging*

All experiments were conducted at the 2-photon core facility (46-6178) at the McGovern Institute for Brain Research (MIBR), MIT. 2-photon excitation fluorescence microscopy (2PEF) for simultaneous structural imaging, glutamate uncaging (2PGU), and calcium imaging was performed using two Ti:sapphire lasers (Mai Tai eHPDS, Spectra Physics), with a custom-built optical setup including an 8x pulse splitter (https://flosfor.github.io/pulse_splitter.pdf).

Optical elements in the excitation and emission pathways were obtained from Thorlabs, or Chroma in the case of some optical filters, unless otherwise specified. Signals were acquired with two GaAsP detectors (H7422-40, Hamamatsu), after wavelength separation. For structural and calcium imaging, cells were visualized with Alexa Fluor 594 and Oregon Green BAPTA-1 included in the internal solution, with 2-photon excitation wavelength respectively at 880 nm or 920 nm. For calcium imaging, line scans (256 repetitions of ~10 μm lines, each with ~1 ms duration at 8.25 μs/pixel) were made across the dendritic branch segment, near the center of the synaptic spine cluster that was activated with 2PGU. The region of interest (ROI) and the background for calculating the fractional change in calcium fluorescence (dF/F) were defined respectively as those points with signal intensities above the 98$^{th}$ percentile or below the 50$^{th}$ percentile from structural imaging during the baseline period prior to 2PGU pulse delivery, with the ROI consisting of continuous points flanked by two segments each containing separate parts of the background. For 2PGU, 5 mM MNI-caged-L-glutamate (Tocris) was dissolved in the recording aCSF (adjusted to 120 mM NaCl from 125 mM to compensate for osmolarity), and delivered with minimal pressure using a puffer pipette of pore diameter ~15-20 μm, positioned < 100 μm away from the synaptic spines at an incident angle of ~70°. 2PGU was achieved with the 2-photon excitation wavelength of 720 nm, quasi-simultaneously at an interval of 0.12 ms between each uncaging spot with a dual galvanometer scanner (Bruker), ~0.1 μm away from the spine head, at an intensity producing somatic uEPSPs with physiological kinetics and amplitudes, typically ~0.1-0.5 mV (corresponding to ~50-150 mW in terms of the optical power of the excitation laser measured past the 63x objective (421480-9900-000, Zeiss), but with continuous full-field illumination instead of short, focalized pulses used for actual 2PGU experiments). A total of 20-35 spines, typically spanning ~30-50 μm along a continuous dendritic branch segment, were activated at a given location.

*Data analysis*

Electrophysiology and calcium imaging data were analyzed and visualized using PVBS (https://github.com/flosfor/pvbs). Structural imaging data were processed using imageJ (https://imagej.net) with the MosaicJ plugin (https://github.com/fiji-BIG/MosaicJ). Plots were made using Prism (GraphPad), custom codes written in MATLAB (Mathworks), and PVBS. Values were expressed as mean ± standard error of the mean (SEM), with n indicating the number of branches, cells, or humans, as stated. The skewness and the excess kurtosis of distributions were denoted respectively by $G_1$ and κ; $G_1$ is the Fisher-Pearson standardized third moment coefficient, and $κ = (μ_4/σ^4) - 3$, where $μ_4$ is the fourth central moment and σ is the standard deviation. Statistical comparisons were made with the Mann-Whitney U test, unless otherwise specified. Spearman's rank correlation coefficient (ρ) was used to evaluate correlations. Linear regressions were calculated based on the least squares method, with the coefficient of determination ($R^2$) from the Pearson correlation coefficient (r) reported. The slopes (m) of the linear regressions were compared using the F-test, including their deviation from zero. Statistical significance was accepted when the associated P value was below 0.05 (* $P < 0.05$; ** $P < 0.01$; *** $P < 0.001$).

**Acknowledgements**

We thank the clinical and research coordination personnel at MGH and BWH for the availability of human brain tissue. We also thank Dr. Suk-Ho Lee for the critical reading of the manuscript. This work was supported by the Y. Eva Tan Fellowship from the K. Lisa Yang and Hock E. Tan Center for Molecular Therapeutics in Neuroscience at MIT (2021-2023, to J. Y.).

**Figures and Legends**

**Fig. 1**

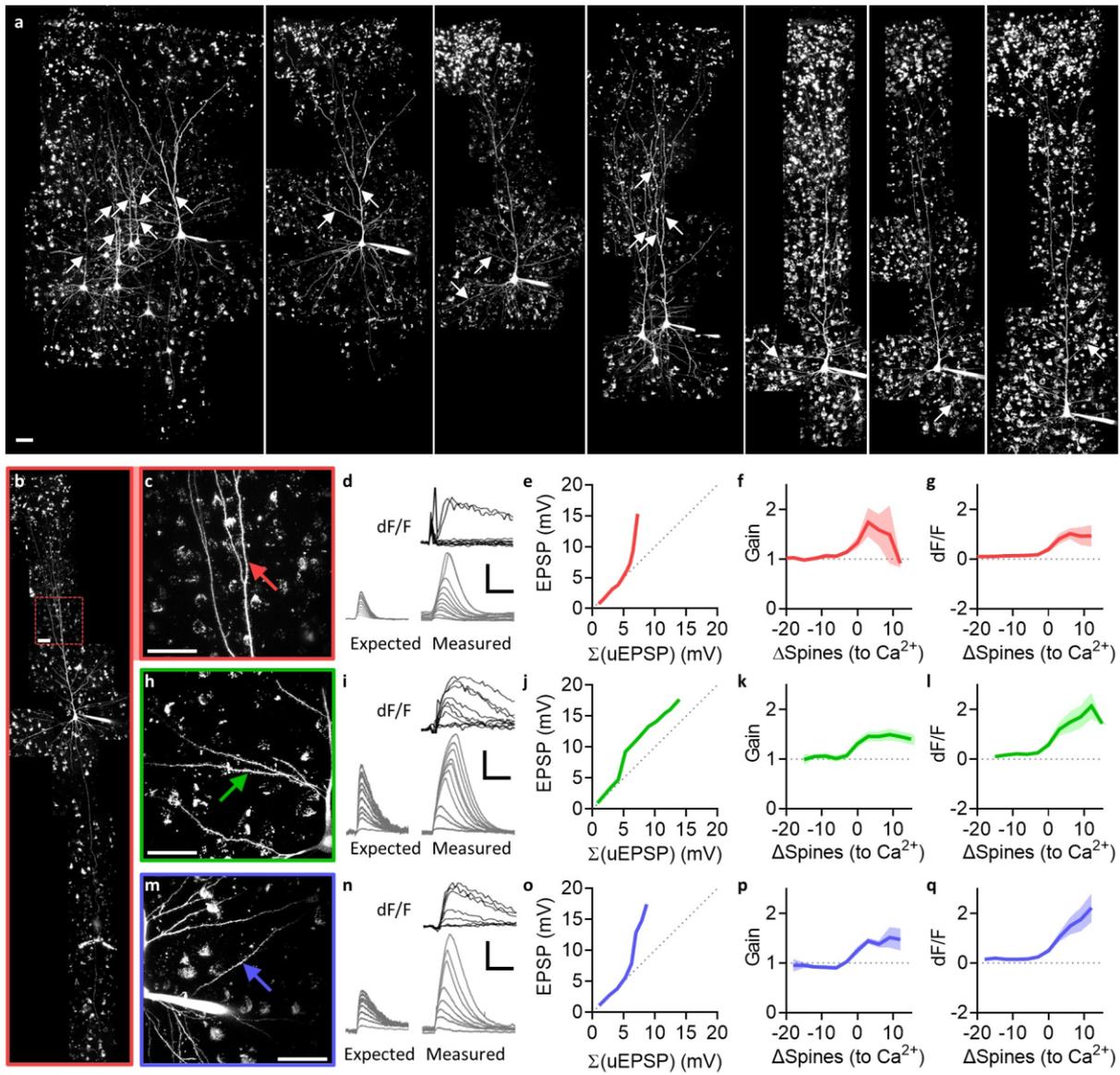

**Figure 1.** Synaptic integration at human neocortical layer 2/3 pyramidal neurons (L2/3 PN). **(a)** Representative examples of human L2/3 PNs. Scale bar, 50 μm. Scale bar definitions are consistent throughout all images in all figures. Arrows indicate the center of synaptic spines that were activated by 2-photon glutamate uncaging (2PGU), which typically spanned approximately 30-50 μm along the length of a dendritic branch segment. The high background fluorescence in the human cortex is caused by lipofuscin aggregates on the neuronal somata; see **Fig. 4** for comparison with the rodent cortex. **(b-c)** Representative example of a 2PGU location at the apical tuft. **(d)** Representative data from the dendrite in panel **c**, with the expected EPSP (left) calculated from the arithmetic sum of unitary EPSPs (uEPSP) recorded from each respective spine, and the measured EPSP (right, bottom) recorded by simultaneous activation of multiple spines, along with the associated fractional change in fluorescence (dF/F) from intracellular calcium imaging at the parent dendrite (right, top). Scale bars, 50 ms, 5 mV, 1.0 dF/F. **(e)** Representative data from the same dendrite as in panels **b-d**, with the measured EPSP plotted against the expected EPSP from the arithmetic sum of uEPSPs. **(f)** Grouped average of synaptic gain from all apical tuft dendrites (n = 24). Gain was defined as the ratio of measured vs. expected EPSP. The number of activated spines were aligned to the nonlinearity threshold at which $Ca^{2+}$ signal was first observed at the synaptic site. **(g)** Grouped average of dF/F, associated with data shown in panel **f**. **(h-l)** Similar to panels **c-g**, but from oblique dendrites (n = 16) branching from the apical trunk. **(m-q)** Similar to panels **c-g** or **h-l**, but from basal dendrites (n = 20) extending from the soma.

**Fig. 2**

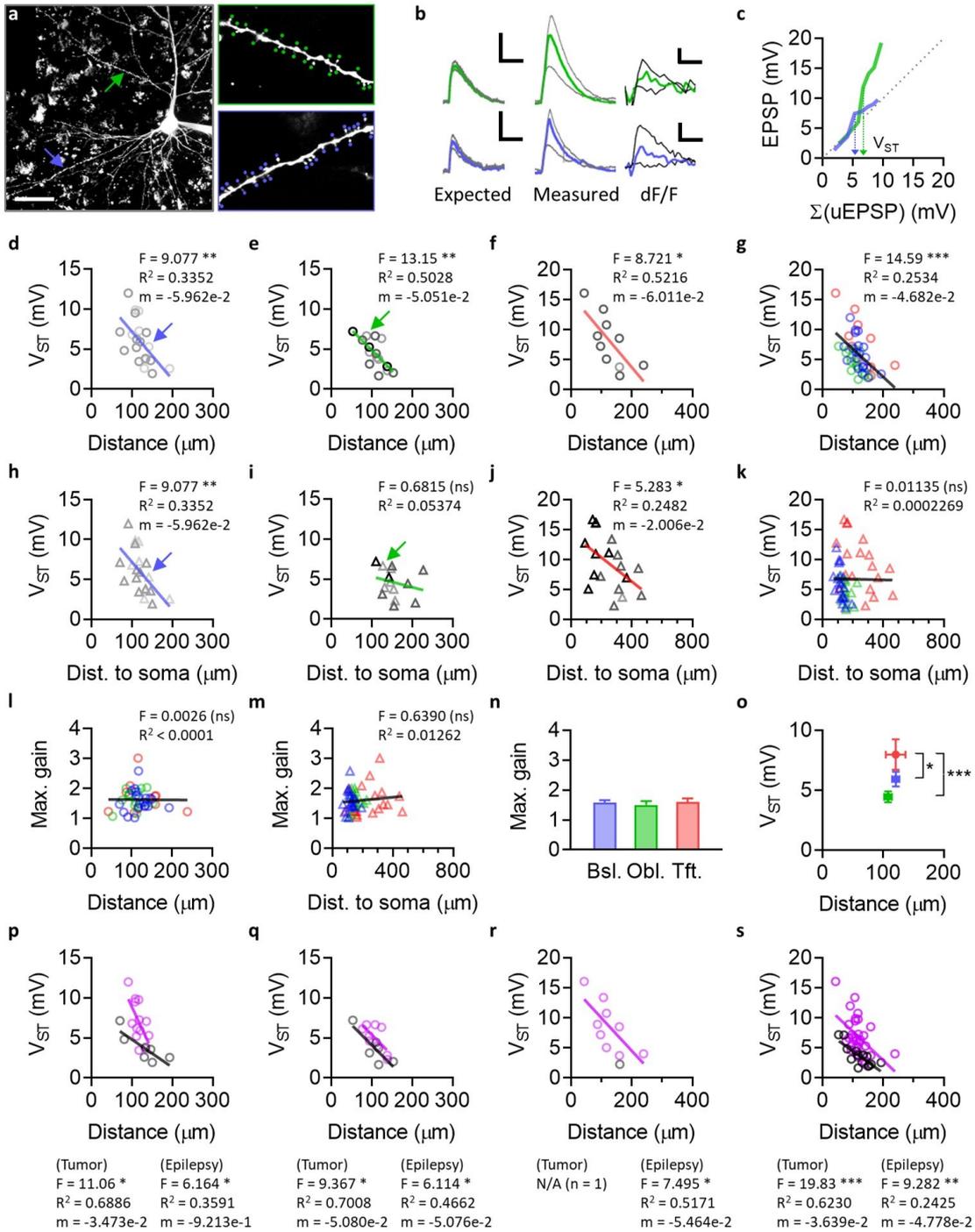

**Figure 2.** Somatic depolarization at supralinearity threshold ($V_{ST}$) is linearly determined by synaptic distance from the apical trunk. **(a)** Representative example of a human L2/3 PN. Two different uncaging locations, one at a basal dendrite (blue) and another at an oblique dendrite (green), are indicated by arrows. Scale bar, 50 μm (left). Dots indicate uncaging spots (right). **(b)** Representative traces of expected and measured EPSP, along with the associated dF/F, from the branches shown in panel **a**. For clarity, only the sweeps at or immediately before and after the nonlinearity threshold are shown. Scale bars, 50 ms, 5 mV, 0.1 (top) or 0.5 (bottom) dF/F. **(c)** Measured EPSP vs. expected EPSP from the sum of uEPSPs, from the same branches as in panels **a-b**. $V_{ST}$ was defined as the expected EPSP at synaptic nonlinearity threshold indicated by the local $Ca^{2+}$ signal. The expected EPSP from the sum of uEPSPs was used for the definition of $V_{ST}$ to determine the correct threshold irrespective of the nonlinear gain. **(d-g)** $V_{ST}$ plotted against synaptic distance. Synaptic distance was defined as the shortest projected path distance along the dendrite from the synapse to the apical trunk. Notably, the linear regressions in panels **d-f** had similar slopes in all groups (F = 0.06380, P = 0.9383), and consequently also in panel **g**. Shades of symbols indicate branch order (darker to lighter, 1st to 4th; throughout panels **d-f**), which was not correlated with $V_{ST}$ in any group. **(d)** $V_{ST}$ vs. synaptic distance at basal dendrites (n = 20). **(e)** $V_{ST}$ vs. synaptic distance at oblique dendrites (n = 15). **(f)** $V_{ST}$ vs. synaptic distance at apical tuft dendritic locations beyond the nexus (n = 10). **(g)** Data shown in panels **d-f**, taken together for comparison. **(h-k)** Similar to panels **d-g**, but plotted against the distance from the synapse to the soma instead of the apical trunk. Panel **h** was reproduced from panel **d** as the path distance from the synapse to the soma or the apical trunk are approximately identical for basal dendritic locations, since the position of the soma was defined as the center of mass of its boundary. In panel **j**, synapses located at the nexus are also included (n = 18), which were excluded from panel **f** due to having zero synaptic distance per definition. **(l)** Maximum gain vs. synaptic distance to the apical trunk. **(m)** Maximum gain vs. distance to soma. **(n)** Maximum gain at basal, oblique, and tuft dendrites. **(o)** Average $V_{ST}$ at basal, oblique, and tuft dendrites. **(p-s)** Same data as in panels **d-g**, but grouped according to tissue origin in terms of tumor (black, n = 46) or epilepsy (purple, n = 45). Neurons from the nonpathological part of the neocortex associated with epilepsy had a tendency for higher $V_{ST}$ at a given synaptic distance, compared to those associated with heterogenous tumor.

**Fig. 3**

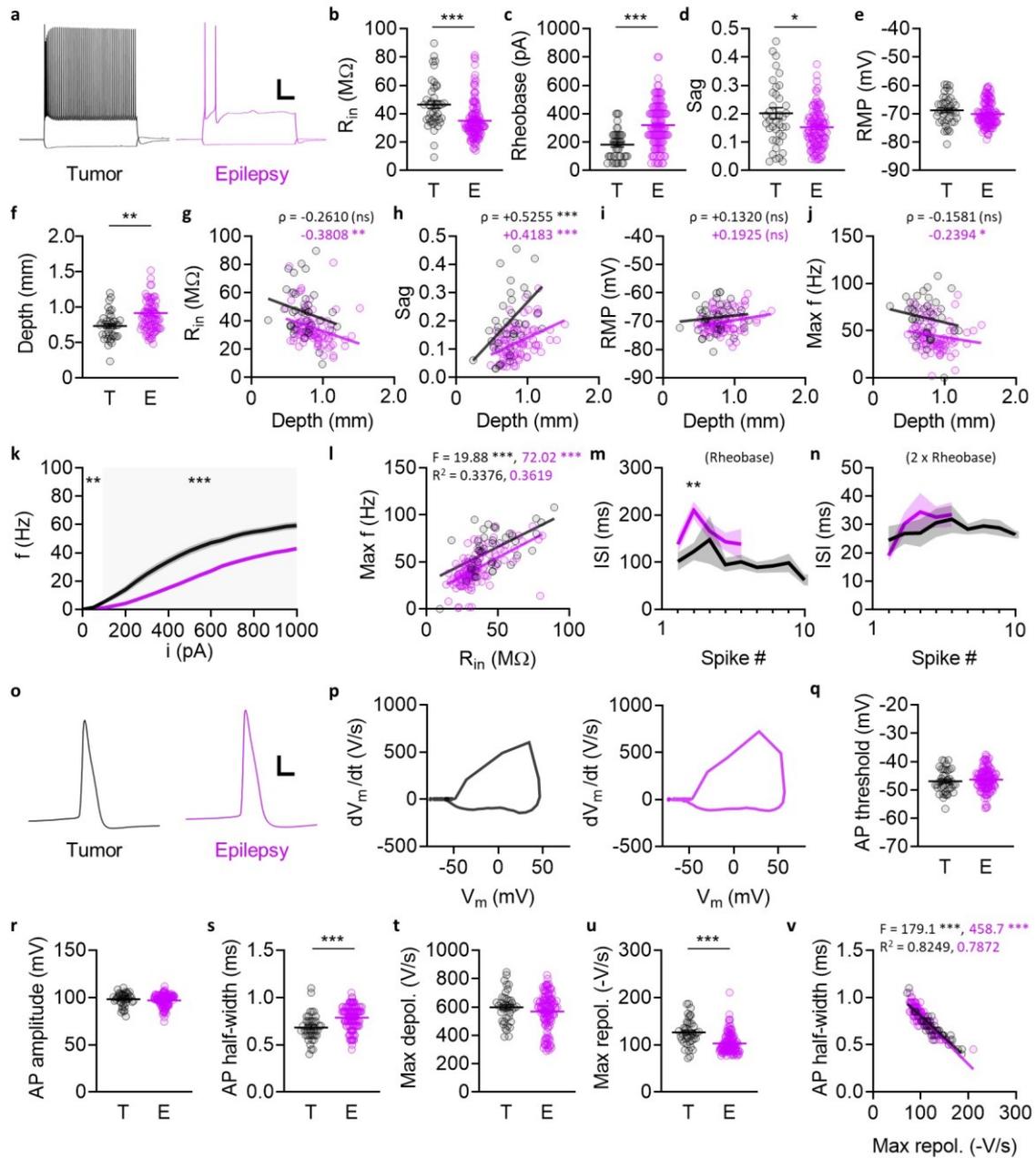

**Figure 3.** Intrinsic membrane properties of human L2/3 PN, from tumor (black) or epilepsy (purple) patients. For more detailed patient and tissue information, see **Table 1**. Note that tissue from both tumor and epilepsy patients originated from parts of the neocortex that were clinically categorized as nonpathological. **(a)** Representative traces of membrane potential responses to somatic step current injection (-250 pA, +1000 pA). Scale bar, 20 mV, 200 ms. The example to the right was taken from the same cell as in **Fig. 1b**. **(b)** Input resistance ($R_{in}$), of L2/3 PNs associated with heterogenous tumor (n = 41) or epilepsy (n = 129). See Methods for $R_{in}$ calculation. **(c)** Rheobase, with current step resolution of 50 pA. **(d)** Sag ratio. See Methods for the definition of sag ratio. **(e)** Resting membrane potential (RMP). **(f)** Cortical depths of L2/3 PNs included in the current study. Cortical depth of a cell was defined as the linear distance between the pial surface and the soma, extrapolated from the straight line connecting the soma and the nexus. The depth of one outlier in the tumor group (235 μm) and its identity as an L2/3 PN was verified with the whole-cell 2-photon image of the cell, accompanied by other L2/3 PNs deeper in the vicinity, all of which had dendritic arbors that were in plane and intact up to the pia mater. **(g-j)** Intrinsic membrane properties, plotted against cortical depth. Linear regressions are presented for visual aid, but correlations were assessed using Spearman's rank correlation coefficient. **(g)** $R_{in}$. **(h)** Sag ratio. **(i)** RMP. **(j)** Peak firing rate. **(k)** Firing rate in response to somatic current injection. L2/3 PN firing rates were significantly lower in epilepsy compared to tumor ($P < 0.001$ each for all i ≥ 100 (pA), $P < 0.01$ for i = 50). **(l)** Peak firing rate, plotted against $R_{in}$. **(m)** Inter-spike interval (ISI) at rheobase. **(n)** ISI at 2*rheobase. **(o)** Representative traces of single action potentials (AP) from human L2/3 PNs. For each cell, the first AP generated at rheobase was taken for analysis. Scale bar, 20 mV, 1 ms. The example to the right was taken from the same cell as in **Fig. 2a**. **(p)** Representative examples of AP waveforms corresponding to the same traces shown in panel **o**, presented as the derivative of the membrane potential with respect to time ($dV_m/dt$) plotted against the membrane potential ($V_m$). **(q)** AP threshold. AP threshold was defined as the $V_m$ at which $dV_m/dt$ reached 10 (V/s). **(r)** AP amplitude (from AP threshold to AP peak). **(s)** AP half-width. **(t)** Maximum rate of depolarization. **(u)** Maximum rate of repolarization. **(v)** AP half-width plotted against the maximum rate of repolarization.

**Fig. 4**

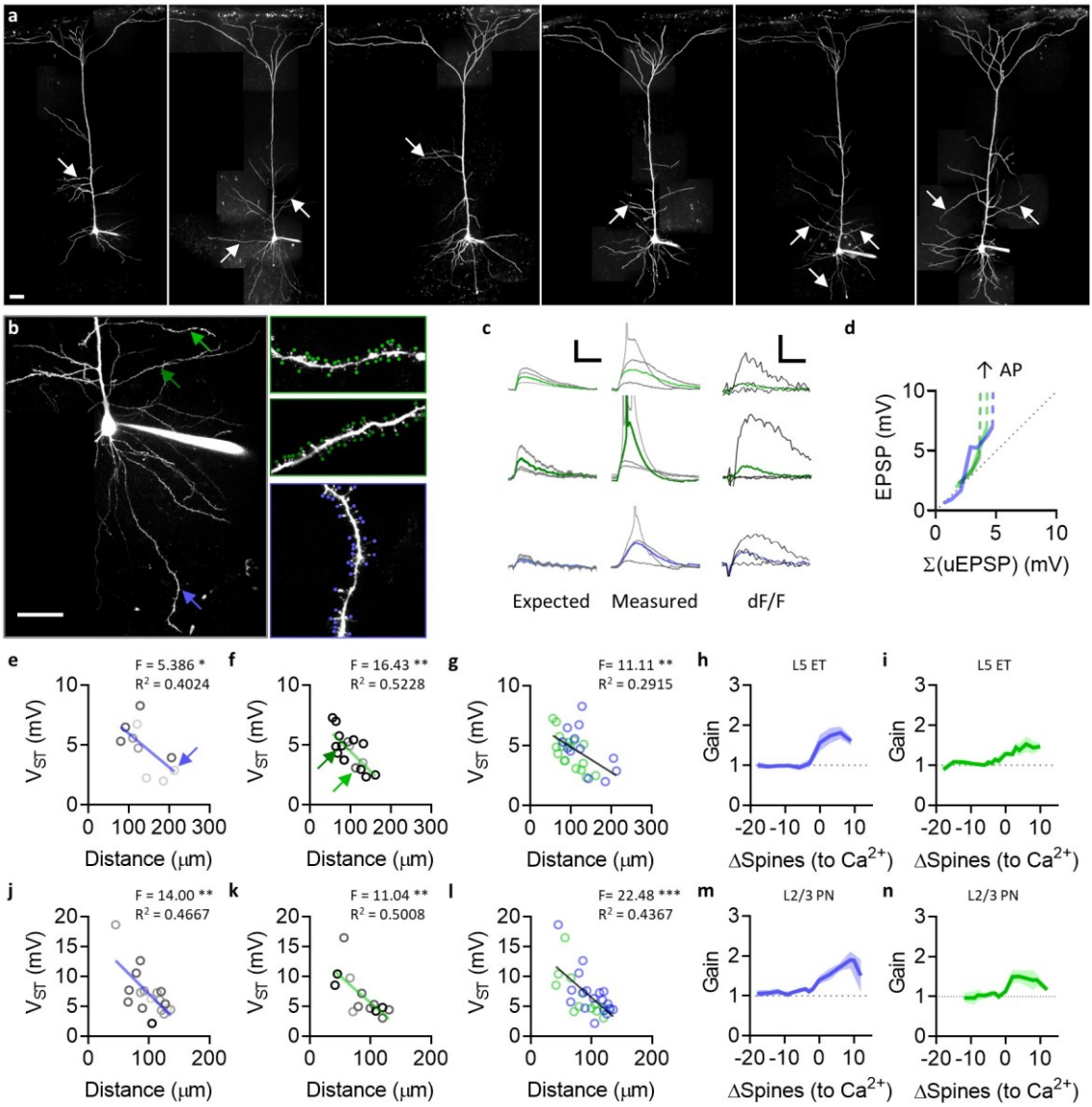

**Figure 4.** Somatic depolarization at supralinearity threshold ($V_{ST}$) is determined by synaptic distance from the apical trunk, analogously for supragranular and infragranular PNs in the rodent cortex in addition to human PNs. **(a)** Representative examples of L5 extratelencephalic (ET) PNs, from the rat temporal association area (TeA). Arrows indicate uncaging locations. Scale bar, 50 μm. Compare with **Fig. 1** for the difference in background fluorescence caused by lipofuscin in the human cortex. **(b)** Representative example of a rat L5 ET PN, with three different uncaging locations; one on a basal dendrite, and two on separate oblique dendrites. Dots indicate uncaging spots. **(c)** Representative traces from synaptic activation in dendrites shown in panel **a**. Scale bars, 50 ms, 5 mV, 0.5 dF/F. **(d)** Measured EPSP vs. expected EPSP from the sum of uEPSPs, from the same dendrites as in panels **b-c**. **(e-g)** $V_{ST}$ vs. synaptic distance as defined in **Fig. 2**, in rat L5 ET. The slopes of the regressions in panels **e** and **f** were not different (F = 0.5381, P = 0.4700). **(e)** Basal dendrites (n = 10). **(f)** Oblique dendrites (n = 17). **(g)** Data shown in panels **e-f**, overlaid for comparison. **(h)** Synaptic gain, at basal dendrites shown in panel **e**. **(i)** Synaptic gain, at oblique dendrites shown in panel **f**. **(j-n)** Similar to panels **e-i**, but from rat L2/3 PNs instead of L5 ET. The slopes of the regressions in panels **j** and **k** were not different (F = 0.1038, P = 0.7498). **(j)** Basal dendrites (n = 18). **(k)** Oblique dendrites (n = 13). **(l)** Data shown in panels **j-k**, overlaid for comparison. **(m)** Synaptic gain, at basal dendrites shown in panel **j**. **(n)** Synaptic gain, at oblique dendrites shown in panel **k**.

**Table 1**

| ID | Age | Sex | Area | Hemi. | Medications (AED) | Diagnosis | n (cells/humans) |
|---|---|---|---|---|---|---|---|
| BA14 | - | - | ATL | - | - | Epilepsy | 6 |
| BA27 | - | - | TL | - | - | Epilepsy | 6 |
| BB05 | - | - | FL | - | - | Epilepsy | 1 |
| BF30 | 34 | F | TL | R | - | Epilepsy | 1 |
| BI30 | - | - | TL | - | - | Epilepsy | 3 |
| BJ04 | - | - | TL | - | - | Epilepsy | 3 |
| BJ06 | - | - | FL | - | - | Epilepsy | 1 |
| BK02 | 55 | M | ATL | R | Le | Tumor | 10 |
| BK05 | 32 | M | TL | L | Cb, Le, O, T | Epilepsy | 2 |
| BK10 | 71 | F | TL | R | - | Tumor | 13 |
| BK18 | 46 | M | FL | L | Ce, O, Ph | Epilepsy | 7 |
| BL06 | 71 | M | TL | R | - | Epilepsy | 13 |
| CC07 | 33 | M | FL | R | B, Lc, Lo | Epilepsy | 12 |
| CC17 | 23 | F | ATL | R | Cn, G, Lm, Le, Lo, T, Z | Epilepsy | 43* |
| CC31 | 24 | M | ATL | R | Lc, Le, O, T | Epilepsy | 1 |
| CG05 | 18 | F | TL | R | Cb, G, Lm, Le, O, T | Epilepsy | 6 |
| CH15 | 57 | F | TL | R | Le | Tumor | 9 |
| CH16 | 24 | M | TL | R | E, G, Lm, Le, O, Pe | Epilepsy | 16 |
| CI08 | 30 | M | TL | L | Lm, Le | Epilepsy | 1 |
| DA31 | 72 | F | TL | R | - | Tumor | 4 |
| DB16 | 31 | M | TL | R | - | Epilepsy | 7 |
| DC15 | 71 | M | TL | R | - | Tumor | 7 |
| | | | | | | (Tumor) | 43/5 |
| | | | | | | (Epilepsy) | 129/17 |
| | | | | | | (Total) | 172/22 |

Abbreviations:
B, brivaracetam; Ce, cenobamate; Cb, clobazam; Cn, clonazepam; E, eslicarbazepine;
G, gabapentin; Lc, lacosamide; Lm, lamotrigine; Le, levetiracetam; Lo, lorazepam;
O, oxcarbazepine; Pe, perampanel; Ph, phenytoin; T, topiramate; Z, zonisamide;
-, not available (not applicable or unknown). Note that only limited information was made available for a subset of cases from the epilepsy group, which accounts for ~11.6% of all recorded cells (age and sex unavailable, n = 20/6 (cells/humans) from a total of 172/22).
* Recorded up to 120 h post-resection (typically ~48 h; see Methods).

**Table 1.** Patient and tissue information.